# Gigahertz directional light modulation with electro-optic metasurfaces

Sam Lin[1], Yixin Chen[1], Taeseung Hwang[1], Anant Upadhyay[2], Ramy Rady[2], David Dolt[2], Samuel Palermo[2], Kamran Entesari[2], Christi Madsen[2], Zi Jing Wong[1,4,*] and Shoufeng Lan[1,2,3,4*]

[1] Department of Materials Science and Engineering, Texas A&M University, College Station, Texas 77843, USA

[2] Department of Electrical and Computer Engineering, Texas A&M University, College Station, Texas 77843, USA

[3] Department of Mechanical Engineering, Texas A&M University, College Station, Texas 77843, USA

[4] Department of Aerospace Engineering, Texas A&M University, College Station, Texas 77843, USA

*Corresponding authors: shoufeng@tamu.edu and zijing@tamu.edu

**Abstract**

Active metasurfaces promise spatiotemporal control over optical wavefronts, but achieving high-speed modulation with pixel-level control has remained an unmet challenge. While local phase control can be achieved with nanoscale optical confinement, such as in plasmonic nanoparticles, the resulting electrode spacings lead to large capacitance, limiting speed. Here, we demonstrate the operation of a gigahertz-tunable metasurface for beam steering through local control of metasurface elements in a plasmonic-organic hybrid architecture. Our device comprises a corrugated metallic slot array engineered to support plasmonic quasi-bound states in the continuum (quasi-BICs). These plasmonic quasi-BICs provide ideal optical confinement and electrical characteristics for integrating organic electro-optic (OEO) materials like JRD1 and have not been previously utilized in optical metasurfaces. We obtain a quasi-static resonance tunability of 0.4 nm/V, which we leverage to steer light between three diffraction orders and achieve an electro-optic bandwidth of ~4 GHz, with the potential for further speed improvements through scaling rules. This work showcases on-chip spatiotemporal control of light at the sub-micrometer and gigahertz level, opening new possibilities for applications in 3D sensing and high-speed spatial light modulation.

**Introduction**

Optical beam steering is a critical capability in emerging sensing and imaging technologies, with applications ranging from LiDAR scanners to augmented reality systems[1–5]. The demand for compact, cost-effective, and high-performance solutions for these technologies has driven research towards nanophotonic devices, particularly metasurfaces, due to their potential for integration with an efficient on-chip footprint. Furthermore, the requirement for dynamic beam steering has naturally pushed the field towards the development of active metasurfaces, prompting the exploration of material platforms like liquid crystals[6–10], phase change materials[11–14], inorganic Pockels materials[15–19], and transparent conducting oxides (TCOs)[20–23].

A critical factor in many applications is operational speed, as scanned depth images require millions of points per second for high angular and temporal resolution[1,24–26]. This mandate for rapid modulation presents a significant challenge in developing active metasurfaces for beam steering applications. For example, liquid crystalline materials struggle to achieve nanosecond-scale response times[6,10,27,28]. Phase change materials have been developed for integrated optics applications but rely on comparatively slow crystallization and heat transfer dynamics[29–31]. Another explored platform is TCOs, such as indium tin oxide, with the capability of carrier-induced unity-order refractive index changes[32]. However, the active regions of TCO-based devices are typically a sub-nanometer carrier accumulation layer. Gap plasmon resonances that form between patterned electrode surfaces can compress the optical field and improve overlap[20,21,23,33]. As a result, typical demonstrations of TCO active metasurfaces have electrode spacing on the order of 5-10 nm[20–23]. Such a constraint increases the intrinsic capacitance, impeding larger metasurfaces for high-speed beam steering under the circuit theorem. Therefore, an alternative approach is urgently needed to alleviate these tradeoffs.

One area of increasing interest is using organic electro-optic (OEO) materials in metasurface architectures. OEO materials have been making steady progress for the past decade, enabling their implementation in various integrated optical modulator architectures[34–39] and more recently, metasurface-based free-space modulators[40–48]. Latest developments have further extended the scope of these materials to visible wavelengths[49]. An inherent property of this material family is the field-driven refractive index change throughout an active volume. This property can potentially increase operational speed by expanding electrode spacing without compromising tunability. Despite this advantage and several high-speed metasurface demonstrations[40–42], a photonic architecture that effectively leverages organic chromophores for phased-array beam steering at high switching rates has, to our knowledge, still been elusive.

In this work, we uncover a pathway towards incorporating OEO chromophore JRD1 into a beam steering metasurface through the physics of plasmonic quasi-bound states in the continuum (quasi-BICs). By modifying an archetypal plasmonic slot waveguide structure and exploiting the electro-optic properties of JRD1[50], we realize a plasmonic-organic hybrid (POH) metasurface for beam steering between three diffraction orders. To demonstrate its bandwidth capability, the modulation of light up to a bandwidth of 4 GHz was measured. Our findings establish design principles for a family of compact phase modulators to exploit new generations of emerging chromophores with high electro-optic coefficients.

**Results**

The design and operation of our POH metasurface relies on creating electric field confinement in both high-frequency optical and low-frequency electrical domains to maximize the electro-optic interaction within the JRD1 chromophore layer. As shown in Fig. 1a, the device architecture employs two input electrodes ($V_1$ and $V_2$) and one ground (GND) electrode to enable dynamic control over reflected light. The fundamental building block of our design is a modified plasmonic slot waveguide, where resonant states are confined between metallic electrodes that serve both optical and electrical functions. The incorporation of JRD1 chromophores (molecular structure shown in Fig. 1a inset) enables efficient conversion of electrical signals to optical phase modulation.

The operational principle of our device is illustrated in Fig. 1b-c. During the initial poling process, an electric field ($E_{\text{poling}}$) aligns the JRD1 chromophores within the plasmonic structure. Subsequently, externally applied electric fields ($E_{\text{applied}}$) can modulate the local refractive index of the chromophore layer. As demonstrated in Fig. 1c, each POH resonator exhibits a $2\pi$ phase change across its resonant spectrum. When the applied field is oriented parallel (red) or antiparallel (blue) to the poling direction, the resonance shifts from its neutral state (purple). By applying different fields to adjacent unit cells, we can generate a phase gradient at the operation frequency (indicated by the green dashed line), enabling beam steering to the +1st diffraction order (m = +1).

To achieve the necessary conditions for electromagnetic confinement and phase control, we engineered the plasmonic structure's geometry. Starting from the basic straight plasmonic slot waveguide, we introduce a sawtooth structural perturbation along the waveguide direction ($\hat{x}$), forcing traveling wave modes to undulate side to side (Fig. 2a), where the perturbation angle $\alpha = 22.5°$ measures the deviation of sawtooth edges away from $\hat{x}$. From the band structure point of view, this modification allows previously traveling wave modes at the X point to fold into the light cone at the $\Gamma$ point due to symmetry reduction[51–53]. To prevent in-plane radiative losses at the metasurface x-edges, we flatten the tips of the metal sawtooth to open a band gap and suppress group velocity ($v_{g,x}$). We further suppress any leakage by increasing the period of the 10 outermost unit cells to create a mirror effect through band gap engineering[54]. The resulting standing wave mode and the band structure are shown in Fig. 2b and Fig. 2c, respectively. Because $\alpha$ characterizes the degree of symmetry breaking and the coupling of radiation to free space, this structure can be considered a new archetype of a quasi-BIC (see Sec. 3 of the Supplementary Information).

Having established the resonator design, we next optimized its functionality as a phased array element. The incorporation of a back reflector constrains radiation to the upward ($+\hat{z}$) direction, allowing each unit cell of the system to be characterized using a one-port resonator model within the framework of coupled mode theory[55]. An effective tunable phased array element requires a $2\pi$ phase shift, which is achievable when the resonator is in an over-coupled state. Consequently, we proceeded to fine-tune the structure to achieve this over-coupled condition (Fig. 2d-f). We start with the perturbed structure with $\alpha = 22.5°$, which induces a radiative Q factor of about ~25. From here, we focused on fine-tuning radiative coupling via the slot width ($w_{slot}$), which corresponds to the inter-electrode distance. Simulated phase spectra show that a total phase change approaches $2\pi$ across the resonance at larger slot widths of 135-170 nm while vanishing for $w_{slot} = 120$ nm, indicative of a transition from the over-coupled regime to an under-coupled regime (Fig. 2d). Correspondingly, simulated amplitude spectra show amplitude dips that are largest for $w_{slot}$ between 120 nm and 135 nm (Fig. 2e). Extracting the effective radiation linewidth and effective absorption linewidth reveals that resonators with $w_{slot} > $ ~125 nm, operate in the over-coupled regime where radiative coupling exceeds absorption loss (Fig. 2f). Thus, we aim to fabricate devices with $w_{slot} = 165$ nm to remain over-coupled while avoiding reduced phase tuning efficiency at larger $w_{slot}$. The spacing between adjacent resonators was chosen to be $a_y = 800$ nm to maximize the steering angle range to about $\pm 75°$, being close to the ideal antenna pitch of $\lambda/2$ (~750 nm in this case), while maintaining acceptably low crosstalk between elements.

We fabricate sets of metasurfaces using standard fabrication processes, as detailed in the methods section. These sets are separately configured for beam steering and high-speed modulation, differing only in their wiring configurations. Each set comprises devices with a range of slot widths. Fig. 3a-c presents SEM and optical microscope images of a device in the beam steering configuration. To verify phase tunability, we measure the reflectivity spectra across a set of devices. As the slot width decreases with increasing device index, we observe that the reflectivity dip initially deepens and then rises again. This behavior indicates a transition from an overcoupled regime through critical coupling to the onset of undercoupling (Fig. 3d). For subsequent active tuning measurements, we pole device 3 with a measured $w_{slot} = 170$ nm as it is closest to critical coupling while remaining in the overcoupled regime, optimizing phase tunability. To assess the tuning efficiency of the poled device, we select a wavelength slightly off-resonance at 1510 nm and track the reflectivity response while applying a quasi-DC triangle wave to the electrodes (Fig. 3e). From this measurement, we extract a modulation efficiency of 0.40 nm/V, highest reported in chromophore infiltrated metasurfaces[40].

Next, we demonstrate the beam steering capability of POH metasurfaces. Our beam steering devices are constructed with two sets of input electrodes and a single winding ground electrode. This configuration allows for the fabrication of a 3-directional beam steering device using only a single layer of lithography. Fig. 4a provides a section view of the metasurface super-cell that illustrates this connectivity scheme. To activate and control the metasurface, we first poled it by setting the poling voltage on the two input electrodes labeled $V_1$ and $V_2$, respectively. During operation, applied voltage induces a refractive index change, shifting the resonant wavelength and thus allowing for control of the reflected phase at a fixed operating frequency. However, directly programming metasurface element reflectivity with a phase-voltage relation produces suboptimal beam steering efficiency due to substantial phase-amplitude variation near critical coupling. To compensate for this, we optimized the applied voltages on $V_1$ and $V_2$ to minimize the side-lobe level (Fig. 4b-c). This was achieved by modeling the metasurface as an evenly spaced array of single-port resonators with spectral responses measured in Figure 3d, and the model predicts a 3:1 side lobe suppression ratio (Fig. 4d-e) largely reflected in our experimental results (Fig. 4f-g).

Having established the device's behavior under quasi-static conditions, we proceeded to verify its performance at high speeds by characterizing the electro-optic response of the POH metasurface architecture up to the gigahertz range. To estimate the device bandwidth, we began by measuring the electrical S-parameters of the set of metasurfaces configured for free-space modulation. To minimize electrical parasitics and enable compatibility with radio-frequency test equipment, standard ground-signal-ground (GSG) pads were included in the lithographic pattern for each terminal (Fig. 5a, inset). Probing device 3 generated a near-perfectly capacitive response with a bandwidth of 3.9 GHz (Fig. 5b). This indicates that the bandwidth limit of our architecture is primarily determined by device capacitance and system impedance. For device 3, we extract an equivalent capacitance of approximately 820 fF. The other devices in this set yielded similar capacitive responses but varying bandwidths ranging from 3.4 to 4.0 GHz (Fig. 5c). We observed a trend of decreasing bandwidth with increasing device index, consistent with the trend of decreasing slot width causing an increase in the device capacitance.

Next, we characterized the electro-optic bandwidth of device 3 by measuring the modulated light with an optical receiver. We observed the expected roll-off in the frequency response, which follows that of the reflection S-parameters, with a measured bandwidth of 3.6 GHz (Fig. 5d). The correspondence between the electrical and electro-optic bandwidths suggests that the device's speed is primarily limited by its electrical characteristics. Finally, we assessed the digital modulation capability of POH metasurfaces to demonstrate its applicability as a universal platform for light modulation in high-speed data transmission applications. We measured eye diagrams at both 3.5 Gb/s and 4 Gb/s (Fig. 5e-f). Oscilloscope trace data was exported and processed offline, where averaging was applied to compensate for the attenuation in the optics system and the photoreceiver noise.

**Discussion**

In this work, we have introduced a POH quasi-BIC metasurface architecture for gigahertz-speed beam steering. This design is fundamentally based on traveling wave modes, which inherently provide an ideal overlap between optical and electrical fields. By leveraging this new scheme for subwavelength electrical control of metasurface elements, we have demonstrated a previously unrealized capability for gigahertz-speed wavefront modulation. A key advantage of our approach lies in its utilization of field-based index modulation, which can be distributed over a relatively large volume. This characteristic allows our platform to overcome the bandwidth limitations typically associated with electro-optic beam steering metasurfaces. Our POH quasi-BIC metasurface will advance the field of dynamic wavefront control, offering new possibilities for high-speed, compact optical beam steering and modulation.

Beyond its performance demonstrated in this work, the POH quasi-BIC metasurface architecture can be extendable to transform diverse photonics applications. While our work used 50-μm devices, the capacitance-limited bandwidth of POH metasurfaces follows a scaling law that allows it to be tuned inversely proportional to the device area. Such a scaling law expands the bandwidth ceiling for many applications. For instance, free-space digital modulation of an optical fiber with a typical mode size of 10 μm may be performed with a downsized free space modulator[35], pushing the bandwidth ceiling to the order of 100 GHz. Furthermore, recent advancements in electro-optic (EO) polymers have extended their functionality into visible wavelengths, paving the way for visible light beam steering applications. This development is particularly significant as it offers an alternative to current visible-light modulation techniques, such as thermo-optic modulation and liquid crystal technology[6,10,56], which are limited by their relatively slow response times. Our work thus opens a path towards high-speed visible light beam steering to revolutionize advanced display technologies (virtual and augmented reality displays), where rapid control of visible light is required for immersive visual experiences.

**Methods**

*Sample fabrication*

POH metasurfaces were fabricated on quartz substrates, and the fabrication workflow is shown in Sec. 1 of the supplementary information. The Au reflector and $SiO_2$ spacer layers were deposited via e-beam evaporation. For the modulator device, the Au reflector was patterned via electron beam lithography (EBL) to be localized beneath the metasurface, minimizing parasitic capacitive coupling between electrodes for high-speed operation. In contrast, the reflector remained unpatterned for the beam steering device. The metasurface patterns and electrode structures were fabricated in a single step of EBL. Bilayer PMMA resist was patterned using a 30 kV electron beam (TESCAN MIRA3 FEG-SEM), with a 5 nm e-beam evaporated Al layer for anti-charging. The Al was removed with a 10:1 dilution of Al etchant (Transcene Type A) prior to development in 3:1 IPA/H$_2$O at 0 °C. The pattern was transferred onto 100 nm of e-beam evaporated Au with 2 nm Ti as an adhesion layer, and liftoff was performed with an overnight immersion in warm acetone. A 2.5 wt% solution of JRD1 (NLM Photonics) in 1,1,2-trichloroethane (Sigma Aldrich) was prepared and spin-coated on top of the device at 4000 RPM, producing a film that overfills the Au gaps by ~100 nm. The deposited chromophore film was baked for 3 h at 65 °C. Finally, a 50 nm SiO$_2$ capping layer was deposited over each metasurface through a physical mask. Poling was performed by applying a voltage of 12 V on a homemade heating stage in a nitrogen enclosure. The temperature was ramped at 10 °C/min to 83 °C, held for 4 minutes, and then allowed to cool.

*Simulations*

Eigenmode, band structure, and reflectance simulations (Fig. 2) were performed in COMSOL Multiphysics 5.6. The refractive indices of SiO$_2$ and JRD1 were assumed to be 1.45 and 1.81, respectively. The optical properties of Au were taken from Johnson and Christy[57].

*Measurement*

Measurements were conducted on an optical setup, and its design is further detailed in Sec. 2 of the supplementary information. Light from a NIR laser source (HP 8168F) was coupled to free space using a GRIN lens fiber collimator (ThorLabs, 50-1550A-FC). A 0.40 NA NIR objective (Mitutoyo, MY20X-824) and a tube lens of focal length 70 mm were used for imaging. During quasi-DC electro-optic measurements, the sample was mounted on the same heating stage used for poling and was contacted via tungsten DC probes on kinematic mounts. A Micro-SWIR 640CSX was employed for NIR imaging, and a photoreceiver (Thorlabs, RXM40AF) was used for intensity measurements.

For high-speed electrical and electro-optic measurements, DC probes were replaced with a pair of GSG probes (Cascade Microtech, I50-A-GSG-125), calibrated with an impedance standard substrate (Cascade Microtech, 101-190). A Rohde & Schwarz ZVA67 vector network analyzer was used for S-parameter and bandwidth measurements. Eye diagrams were collected on a 13 GHz oscilloscope (Agilent, DSA91304A) and exported for offline post-processing. Raw oscilloscope trace data was averaged across 2048 patterns to compensate for signal attenuation dominated by the NIR objective, beam splitter, and low fiber coupling efficiency in the optical system. Additionally, a bit error rate tester (Centellax, PCB12500) was used to generate a pseudorandom bit sequence of length $2^{10} - 1$.

**Fig. 1: Plasmonic-organic hybrid metasurfaces for beam steering.**

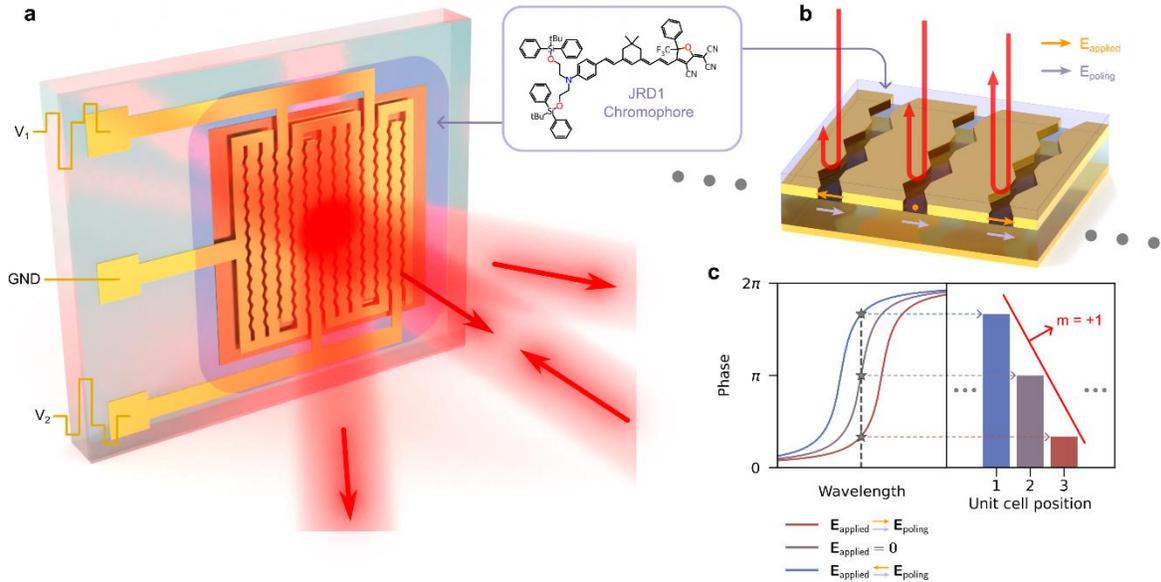

**a,** Schematic overview for a plasmonic-organic hybrid (POH) metasurface. The pictured device is configured with 2 input electrodes, $V_1$ and $V_2$, and a ground (GND) electrode, which together enables the dynamic modulation of the direction of reflected light at gigahertz speed. JRD1, depicted in purple with its molecular structure pictured in the inset, transfers electrical signals into the optical domain. **b,** A conceptual schematic for phase array modulation across POH metasurface unit cells. During poling, the poling electric field ($E_{poling}$) orients JRD1 chromophores within the resonant plasmonic structure. During operation, externally applied electric fields ($E_{applied}$) tune individual resonances via field-induced refractive index change. **c,** A graphical summary of phase array mechanism. Left panel: POH metasurface resonators undergo a $2\pi$ phase change across the resonant spectrum. An applied field parallel to (red) and antiparallel to (blue) the poling field shifts the resonance from the neutral state (purple). Right panel: Appling different fields across adjacent unit cells can generate a phase gradient at the operation frequency (gray dashed line on left panel), thus steering the beam to +1$^{st}$ order (m = +1).

**Fig. 2: Simulation and design of plasmonic-organic hybrid metasurfaces.**

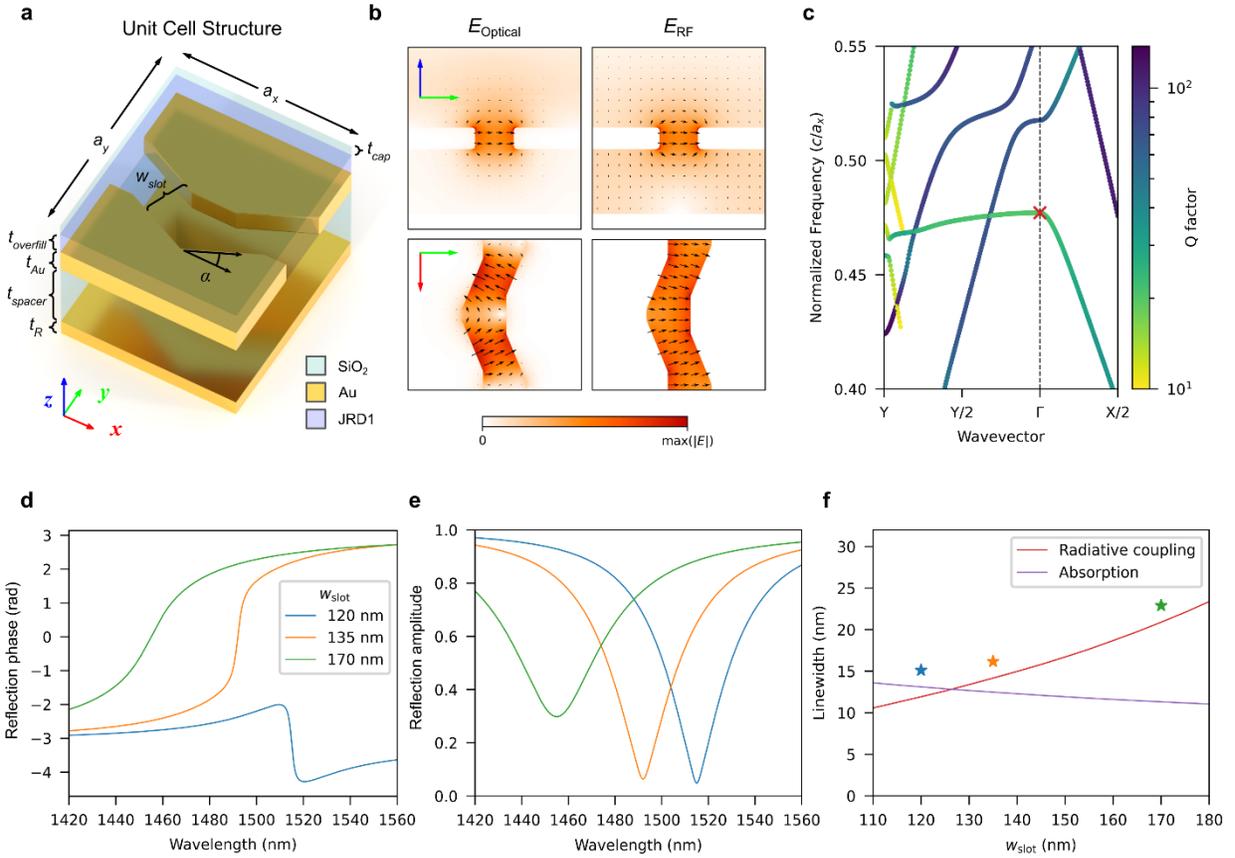

**a,** A perspective view of a unit cell for the POH metasurface. The structural motif is that of a plasmonic slot waveguide array with a sawtooth structural modulation along $x$ with period $a_x$ = 700 nm, separated from a Au reflector by a SiO$_2$ layer, with $t_{spacer}$ = 300 nm. Adjacent waveguides are spaced at a period of $a_y$ = 800 nm. The assembly is JRD1 spin-coated with an oxide capping layer. **b,** Color plots of optical and DC electric field distribution. Left: In the optical domain, POH metasurface resonances inherit the extreme optical confinement of plasmonic slot waveguides. Right: applied electric fields in the DC to RF domain, both spatially coincides and aligns with the optical field which maximizes tuning strength. The top and bottom cross sections are respectively the y-z planes at distance $a_x/4$ from the unit cell boundary in **a** and the midplane of the patterned gold layer. **c,** The band structure. The red cross at Γ marks the mode in the left panels of **b**, generated by folding travelling modes from X into the light cone through a symmetry breaking sawtooth modulation. Flattening sawtooth edges expands the partial band gap for spatial mode confinement along x. The reduced dispersion along Γ-Y indicates limited crosstalk between adjacent unit cells due to plasmonic confinement. **d-e,** Simulated reflection phase (**d**) and amplitude (**e**) spectra show the transition from over-coupled (green and orange) to undercoupled (blue) regime. **f,** The effective linewidths of radiative coupling and the absorption, extracted from the reflection phase and amplitude data. Effective linewidths are measures of the respective decay rates and their sum is the total observed linewidth. Stars indicate the data in **d** and **e**. The 135 nm is optimal for maximal phase modulation in the over-coupled regime with a 2π phase shift across the resonance.

**Fig. 3: Optical and electrical characterizations of fabricated devices.**

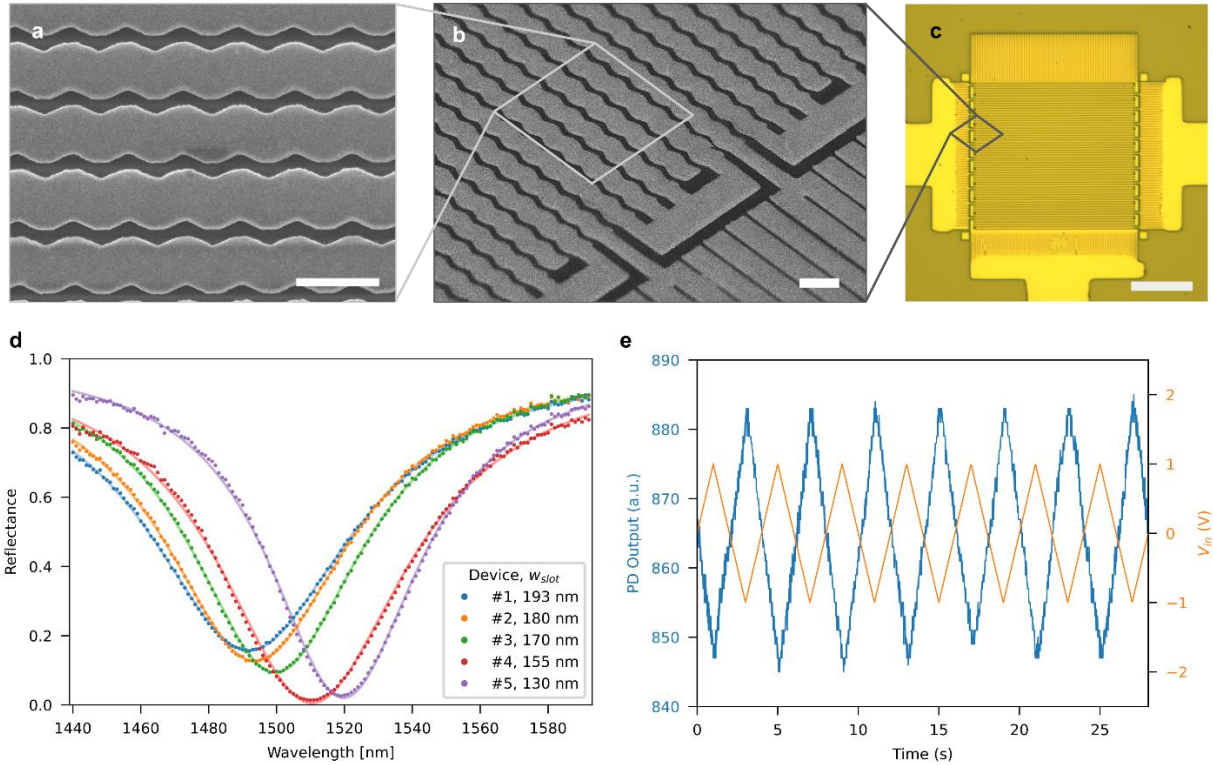

**a-c,** Images of the fabricated device across several length scales. **a,** A scanning electron microscope (SEM) image of the fabricated device, showing the spatial arrangement of unit cells. **b,** A zoomed out SEM view of showing electrode connectivity at the metasurface edges. **c,** A full-device optical micrograph. The scale bars in **a-c** are 1 μm, 1 μm, and 20 μm, respectively. **d,** Measured optical reflectance spectra for five devices fabricated with monotonically decreasing gap size. Light colored curves fit these data to a single-port resonator reflection spectrum. **e,** Quasi-DC electrical modulation of device 3 at 1510 nm wavelength using a 2 V peak-to-peak triangle wave generates optical modulation.

**Fig. 4: Beam steering with electrical control.**

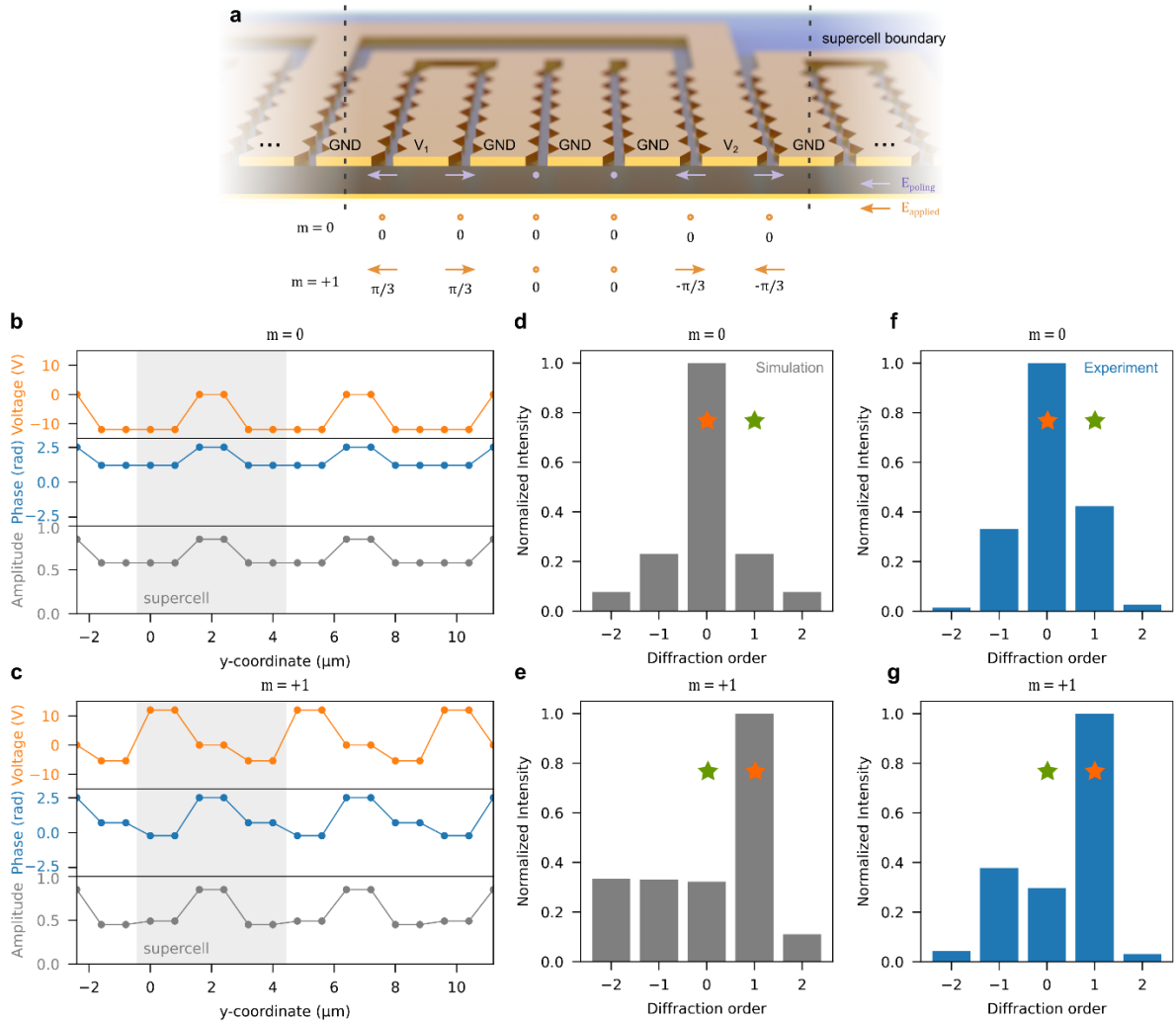

**a,** A graphical schematic of the tuning configuration in a three-electrode beam steering metasurface. The ground electrode nanostructures achieve continuity across the metasurface by connecting at the metasurface edges. During poling, both signal electrodes are held at the poling voltage, whereas during operation they operate independently to generate different directions. The ideal phase and applied electric field direction for beam steering to the $0^{th}$ and $+1^{st}$ diffraction order is listed below the metasurface graphic. **b-c,** Optimized distribution of applied voltages across the metasurface for beam steering into the $0^{th}$ and $+1^{st}$ diffraction orders, along with the resultant reflection phase and amplitude distributions. **d-e,** Simulation results of diffraction order intensity from configurations shown in **b-c**. **f-g,** Experimentally measured beam steering by switching the diffraction from $0^{th}$-order to $1^{st}$-order. In **d-g**, orange-colored stars indicate desired beam steering angles and green colored stars represent unwanted diffraction orders.

**Fig. 5: Gigahertz electro-optic responses of POH metasurfaces.**

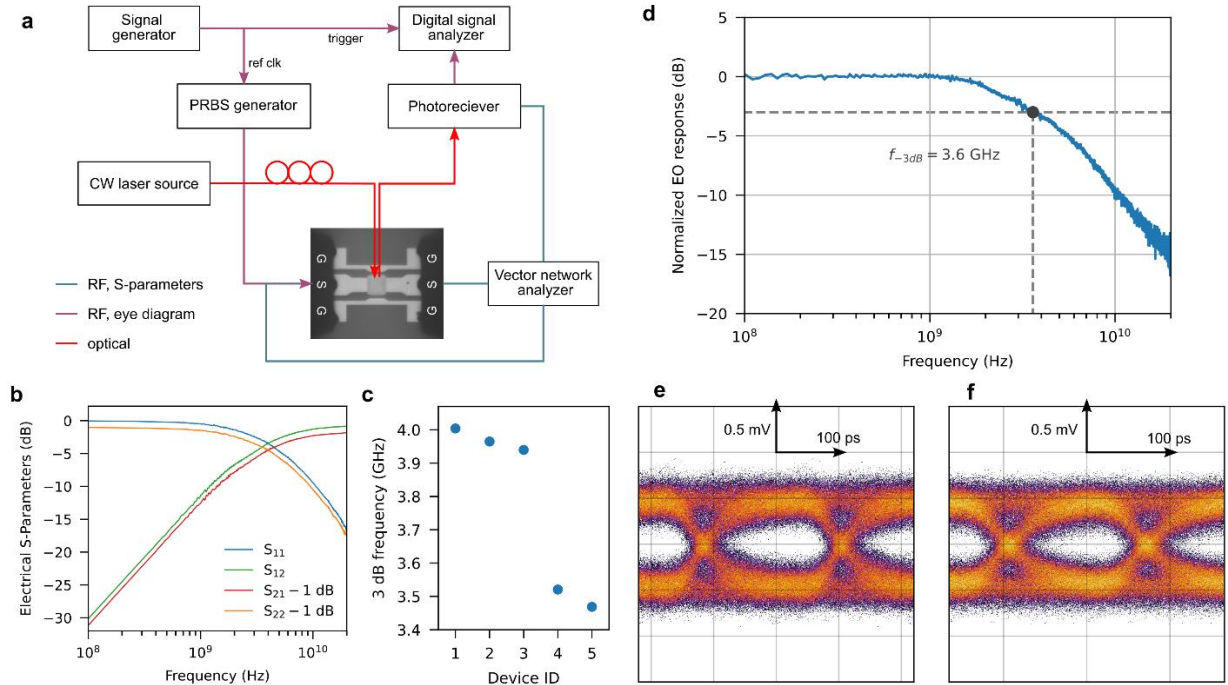

**a,** High-speed measurement setup for a plasmonic slot metasurface in modulator configuration. Teal-colored connections are for measuring S-parameters and purple-colored for acquiring eye diagrams. The inset image shows the landing of two ground-signal-ground (GSG) probes on the device. PRBS: pseudorandom bit sequence; CW: continuous-wave; ref clk: reference clock. **b,** Electrical S-parameters of the device, which show a pure capacitive response. $S_{2x}$ are offset by 1 dB for visibility. **c,** The 3 dB frequency decreases with device index, consistent with an increasing device capacitance as gap size decreases. **d,** The 3 dB rolloff in the electro-optic response of the measured device 3 is 3.6 GHz, which differs slightly from the purely electrical S-parameter result (~3.9 GHz). **e-f,** Eye diagrams measured at 3.5 Gb/s and 4 Gb/s, respectively.